# Interseções entre a Física e os saberes da tradição ceramista

Samuel Antonio Silva do Rosario[1]   Carlos Aldemir Farias da Silva[2]

**Resumo**

A crescente interação entre os saberes da tradição e as ciências formais emerge como um relevante campo de investigação. Nessa perspectiva, o objetivo deste artigo é investigar as interseções entre a Física e os saberes da tradição nas práticas ceramistas da Vila Cuera, em Bragança, estado do Pará. Utilizando uma abordagem etnográfica, complementada por análise qualitativa, percebemos que os ceramistas se valem da intuição sensível e, com maestria, aplicam princípios termodinâmicos na produção de cerâmicas artesanais. Os resultados realçam o valor da Etnofísica, sublinhando que, muito antes da estruturação formal da ciência como a conhecemos, as sociedades tradicionais já praticavam e entendiam fenômenos naturais em sintonia com os princípios físicos contemporâneos. Conclui-se que há um vínculo inerente entre a Física e os saberes da tradição no que se refere à prática ceramista tradicional, fundamentado em observações empíricas e numa sabedoria ancestral que dialoga com a ciência em uma relação de complementaridade. O estudo evidencia a importância de valorizar e reconhecer os saberes das populações tradicionais ao destacar o saber/fazer das culturas e da ciência institucionalizada.

**Palavras-chave:** Física, Saberes da Tradição, Cerâmica, Etnofísica, Termodinâmica.

## Intersections between Physics and the knowledge of the ceramics' tradition

**Abstract**

The growing interaction between traditional knowledge and formal sciences emerges as a relevant field of research. In this perspective, the aim of this article is to investigate the intersections between Physics and traditional knowledge in the ceramics' practices of Vila Cuera, in Bragança, state of Pará. Using an ethnographic approach, complemented by qualitative analysis, we realize that ceramists use their sensitive intuition and, masterfully, apply thermodynamic principles in the production of handmade ceramics. The results highlight the value of Ethnophysics, underlining that, long before the formal structuring of science as we know it, traditional societies already practiced and understood natural phenomena in tune with contemporary physical principles. We conclude that there is an inherent link between Physics and traditional knowledge with regard to the traditional ceramics' practice, grounded on empirical observations and ancestral wisdom, that dialogues with science in a complementary relationship. The study highlights the importance of valuing and recognizing the knowledge of traditional populations by highlighting the knowledge/doings of cultures and institutionalized science.

## Intersecciones entre la Física y los saberes de la tradición ceramista

**Resumen**

La creciente interacción entre los saberes tradicionales y las ciencias formales emerge como un relevante campo de investigación. Desde esta perspectiva, el objetivo de este artículo es investigar las intersecciones entre la Física y los saberes tradicionales en las prácticas ceramistas de Vila Cuera, en Bragança, estado de Pará. Utilizando un enfoque etnográfico, complementado con un análisis cualitativo, percibimos que los ceramistas utilizan su intuición sensible y, con maestría, aplican los principios termodinámicos en la producción de cerámicas artesanales. Los resultados destacan el valor de la Etnofísica, subrayando que, mucho antes de la estructuración formal de la ciencia como la conocemos, las sociedades tradicionales ya practicaban y comprendían los fenómenos naturales en sintonía con los principios físicos contemporáneos. Se concluye que existe un vínculo inherente entre la Física y los saberes tradicionales en respecto a la práctica ceramista tradicional, basado en observaciones empíricas y en una sabiduría ancestral, que dialoga con la ciencia en una relación complementaria. El estudio destaca la importancia de valorar y reconocer los saberes de las poblaciones tradicionales poniendo de relieve los saberes/haceres de las culturas y de la ciencia institucionalizada.
**Palabras clave:** Física, Saberes Tradicionales, Cerámica, Etnofísica, Termodinámica.

---

[1] Instituto Federal do Pará (IFPA), Marabá, Pará, Brasil. E-mail: samuel.rosario@ifpa.edu.br
[2] Universidade Federal do Pará (UFPA), Belém, Pará, Brasil. E-mail: carlosfarias1@gmail.com

# INTRODUÇÃO

A Física é uma das ciências fundamentais da era moderna e tem suas origens intimamente ligadas às práticas e tradições das sociedades ao longo da história humana. Não é apenas uma coletânea de teorias, equações e leis matemáticas, mas, sim, em sua essência, uma manifestação cultural em resposta aos desafios e curiosidades que surgiram na interação das sociedades humanas com o universo e com os fenômenos naturais.

Estudos etnográficos revelam que comunidades tradicionais possuem um vasto repositório de conhecimentos, refletindo uma compreensão profunda de seu entorno e dos fenômenos naturais. Esses saberes não apenas revelam as estruturas sociais e as interações ambientais dessas comunidades, mas também fornecem uma base valiosa para a construção de conceitos que atualmente compõem o repertório da Física, uma vez que o cotidiano das comunidades tradicionais fornece as bases para compreender parte da existência humana desde sempre (ALMEIDA, 2017).

Na contemporaneidade, cabe ressaltar que a Física tornou-se uma disciplina distante do dia a dia para muitos estudantes. Esse afastamento impede que percebam como a Física se faz presente em nossas vidas, pois o desenvolvimento da Física está, em grande parte, intrinsecamente conectado aos contextos socioculturais que nos circundam.

Nesse cenário, surgem estudos voltados para a interação entre os saberes da tradição e os princípios físicos, notavelmente, dentro da Etnofísica – um campo dedicado a abordar a confluência entre a Física acadêmica e as tradições das culturas. Enquanto alguns estudos enfatizam a relação entre o ensino de Física e a cultura dos grupos sociais analisados, outros, adotam métodos etnográficos, oriundos da Antropologia para fomentar o diálogo entre a Física e os saberes de distintas comunidades (ROSARIO; SILVA, 2020a). Este artigo alinha-se à última abordagem.

Na análise de práticas socioculturais das comunidades tradicionais, identifica-se uma "memória biocultural", conforme pontuam Toledo e Barrera-Bassols (2015). Eles enfatizam que essas comunidades, em especial as indígenas, têm como base um legado de princípios relacionados ao funcionamento da natureza. São ensinamentos sobre a convivência sustentável em relação à diversidade biológica e à diversidade cultural como axiomas enraizados em contextos geográficos distintos.

Sociedades imersas em ambientes naturais utilizam uma metarracionalidade para compreender e explicar o mundo, uma ciência sensível e primordial, conforme Lévi-Strauss (2012), pois o saber tradicional se pauta pela "lógica do sensível", que valoriza a experiência direta e a percepção sensorial. Em contraste a essa lógica, o paradigma científico inclina-se para uma abordagem teórica com foco no rigor metodológico.

Paralelamente ao saber científico, comunidades tradicionais desenvolveram ao longo do tempo conhecimentos variados, direcionados à resolução de questões materiais e práticas (ALMEIDA, 2017; MENDES; FARIAS, 2014). Mesmo que façam uso dos mesmos atributos cognitivos universais, esses dois modos de conhecer os fenômenos do mundo (tradicional e científico) são distintos e se pautam por diferentes lógicas de raciocínio: uma centrada e próxima da lógica do sensível, a outra mais distante desta (LÉVI-STRAUSS, 2012).

Sob essa ótica, a produção do conhecimento ocorre em diferentes momentos no espaço e no tempo. Por essa razão, os variados povos espalhados pelo planeta desenvolveram suas próprias formas de compreender e de se relacionar com os fenômenos naturais. Em nossa perspectiva, essas compreensões podem dialogar e contribuir para a Educação em Ciências, em especial neste artigo, a partir da termodinâmica.

Nossa abordagem visa as inter-relações entre os saberes físicos e os saberes da tradição, oriundos das experiências singulares de comunidades tradicionais (ROSARIO, 2023). Reconhecemos que essas formas de saberes interagem e se amalgamam, dando origem a um conhecimento integrado e global, como os saberes da tradição, expressos por indivíduos com expertise específica em diversas culturas globais.

Neste artigo, ao examinar o processo de modelagem da argila e da queima da cerâmica na comunidade Vila Cuera, em Bragança, estado do Pará, aspiramos a estabelecer conexões entre os saberes da tradição e conceitos da Física, especificamente na termodinâmica. Esse esforço visa enriquecer a compreensão desses processos e contribuir para uma abordagem abrangente da Educação em Ciências e Matemáticas.

Assim, temos como objetivo elucidar as possíveis conexões e interseções entre a Física e os saberes da tradição nas práticas ceramistas da Vila Cuera. Destaca-se que, embora possam existir divergências superficiais, uma abordagem aprofundada revela um entrelaçamento intrínseco entre a ciência contemporânea e as tradições da arte de produzir cerâmica, oferecendo perspectivas enriquecedoras para ambas as esferas.

## MÉTODO DE PESQUISA

Para elucidar as interseções entre a Física e os saberes da tradição, relacionadas à produção cerâmica da Vila Cuera, adotou-se uma abordagem de pesquisa qualitativa dada a sua reconhecida capacidade de interação profunda com a comunidade, permitindo uma visão abrangente e fundamentada do fenômeno estudado, conforme destaca Latour (2006).

Na pesquisa de campo, adotou-se a abordagem etnográfica para descrição e interpretação dos dados coletados, seguindo as diretrizes propostas por Oliveira (2016) e Latour (2006). Esse método permitiu compreender a complexidade e a especificidade do conhecimento relacionado à Física na produção de cerâmicas tradicionais. Para complementar o estudo empírico em campo, realizou-se um levantamento de trabalhos próximos ao tema tratado neste artigo no Catálogo de Teses e Dissertações da Coordenação de Aperfeiçoamento de Pessoal de Nível Superior (Capes), no período de 2005 a 2022.

Este artigo é um fragmento de um projeto de pesquisa mais amplo, que teve início em 2016 e continua até o presente momento (ROSARIO, 2018, 2023). Em todas as etapas da investigação solicitamos a autorização da participação dos artesãos por meio da assinatura do Termo de Consentimento Livre e Esclarecido, mantendo, desse modo, os cuidados éticos da pesquisa com pessoas. O documento esclarecia os objetivos, a metodologia e solicitava, ainda, a autorização para o uso de imagens e informações dos dois artesãos que participaram da pesquisa de campo.

A investigação privilegiou quatro técnicas. A primeira foi a observação participante, em conformidade com Oliveira (2016) e Faermam (2014), que revelou-se uma ferramenta valiosa ao possibilitar uma interação genuína com os ceramistas e a comunidade, fornecendo insights sobre o vasto acervo de conhecimentos tradicionais relacionados à Física, envolvidos na produção cerâmica.

Em seguida, realizamos entrevistas semiestruturadas em diferentes etapas da investigação (OLIVEIRA, 2016). O conjunto das entrevistas permitiram examinar a compreensão própria da comunidade, revelando suas práticas socioculturais, narrativas e perspectivas de compreensão dos fenômenos do mundo, em especial, em relação à produção oleira. O terceiro e o quarto instrumentos de coleta de dados empregados foram os registros fotográficos (ROSARIO; SILVA, 2020b, 2023a) e as gravações de vídeos das práticas dos ceramistas (ROSARIO *et al.*, 2020).

O conjunto desses quatro instrumentos proporcionou uma compreensão abrangente da interação entre os saberes da tradição e o conhecimento científico, focando nos saberes físicos que emergem da prática ceramista tradicional.

## ETNOFÍSICA: UMA BREVE REVISÃO DE LITERATURA

Na investigação, buscamos estabelecer pontes entre o ensino da Física e os saberes da tradição. Ao longo dos anos dedicados a esse estudo (ROSARIO; SILVA, 2020a, 2023a), realizamos uma revisão abrangente de trabalhos que abordam a intersecção entre a Física e as práticas socioculturais. Essa revisão incluiu o Catálogo de Teses e Dissertações da Capes e outras plataformas científicas relevantes. O foco foi identificar investigações que refletissem a proposta central do trabalho: a confluência entre saberes tradicionais e acadêmicos na esfera da Física.

Em eventos centrados na pesquisa e ensino da Física, não é raro encontrar estudos rotulados sob o termo "Etnofísica". Esse termo engloba investigações que correlacionam a Física a práticas socioculturais e a saberes tradicionais. Portanto, adotamos "Etnofísica" como palavra-chave primária em nossa revisão bibliográfica, buscando por pesquisas anteriores no cenário brasileiro. Nossa abordagem neste artigo envolve diálogos que abrangem os múltiplos aspectos dos saberes da tradição, incluindo aprendizado comunitário e experiências sensoriais na produção de cerâmicas tradicionais.

Durante nossa revisão no catálogo anteriormente citado, identificamos nove dissertações de mestrado. Vale observar que, até o ano de 2022, não encontramos teses de doutoramento exclusivamente focadas nessa temática, indicando a novidade e o potencial de expansão do campo no cenário nacional. No panorama acadêmico da Física, diversos pesquisadores têm direcionado seus esforços para entender a intersecção entre o conhecimento científico e o saber tradicional de diferentes grupos socioculturais. Essa integração é ressaltada por meio dos trabalhos a seguir, que abordam como diferentes culturas interpretam e interagem com os fenômenos físicos em seus cotidianos.

Anacleto (2007) foi pioneira em investigar essa confluência, focando no cultivo de arroz por trabalhadores rurais em Capivari do Sul, Rio Grande do Sul. Através de uma abordagem etnográfica, ela revelou que, embora os trabalhadores não estivessem familiarizados com o jargão científico, incorporavam princípios da Física em suas práticas. Inspirando-se na Etnomatemática, a autora argumenta que a Etnofísica pode servir como ponte entre a ciência formal e o entendimento popular, tornando a ciência mais acessível e relevante para a população em geral.

Em um enfoque similar, Corrêa (2016) examinou a produção de farinha de mandioca em Pinheiro, estado do Maranhão. Ela identificou aplicações pedagógicas da Etnofísica no ensino das leis de força, apontando para a possibilidade de um ensino de Física mais contextualizado e enriquecido pelas experiências locais.

A interface da Física com a educação indígena foi tema da investigação de Silva (2016), que coletou percepções de alunos indígenas sobre fenômenos, como temperatura e umidade. Ele concluiu que integrar o conhecimento tradicional ao científico pode enriquecer o ensino de ciências nas comunidades indígenas.

A gastronomia também foi objeto de estudo sob a lente da Etnofísica. Silva (2017) investigou essa relação na cidade de João Pinheiro, Minas Gerais, defendendo que os procedimentos físicos cotidianos podem ser valiosos no ensino escolar, tornando a aprendizagem mais significativa.

Rosario (2018), um dos autores do presente artigo, investigou as intersecções entre a Matemática e a Física na produção de cerâmica em Bragança, Pará. Utilizando a Etnografia como seu principal método, abordou as conexões entre os conhecimentos acadêmicos e as práticas socioculturais amazônicas. Seus achados sugerem que tais conexões enriquecem o ensino das Ciências e promovem uma compreensão dos saberes inerentes às diversas comunidades da Amazônia. O autor aborda a Etnofísica, salientando seu papel na análise dos saberes e habilidades de diferentes grupos socioculturais. Essa abordagem se concentra na contextualização dos fenômenos físicos estudados, com o objetivo de revitalizar e valorizar os saberes etnofísicos em sintonia com a perspectiva da Física científica.

Oliveira (2018) explorou como a Etnofísica se manifesta no cotidiano escolar. Utilizando métodos criativos como contação de histórias e desenhos, revelou a importância de relacionar o conhecimento científico ao dia a dia dos alunos.

Almeida (2019) investigou os saberes cosmológicos do povo Paiter-Suruí, evidenciando sua relevância para o ensino da Física. Esses conhecimentos, originados na terra indígena Sete de Setembro, entre Rondônia e Mato Grosso, ilustram a percepção única dos suruís sobre a natureza e o universo. Almeida defende que a Etnofísica pode ser uma ponte entre o conhecimento científico e os saberes tradicionais, enriquecendo a educação e valorizando as perspectivas indígenas na sociedade contemporânea.

Abreu (2021) propôs uma abordagem da Etnofísica para ensinar cinemática na Educação de Jovens e Adultos em Abaetetuba, Pará, utilizando uma sequência didática centrada no regionalismo amazônico. Após a aplicação, houve maior interação entre alunos e educador. Os alunos demonstraram mais interesse e autonomia ao relacionarem a Física com seus conhecimentos regionais. Por fim, em um estudo paralelo no mesmo ano e município, Feio (2021) empregou a produção de farinha de mandioca como estratégia da Etnofísica para ensinar termodinâmica a alunos do ensino médio, concluindo que atividades de campo regionalizadas potencializam o interesse dos estudantes em conceitos físicos.

A Etnofísica, descrita nos nove trabalhos de mestrado, é resultante da intersecção dos saberes tradicionais e os conhecimentos científicos da Física e enfatiza como diferentes culturas interpretam e interagem com fenômenos naturais. Ao integrar os saberes tradicionais à ciência moderna, a Etnofísica enriquece o ensino e a compreensão da Física. Essa abor-dagem não só valoriza a pluralidade cultural, mas também promove uma educação científica mais inclusiva e contextualizada. Ao analisar os trabalhos aqui mencionados, nota-se a variedade de métodos adotados, desde etnografias até sequências didáticas, destacando a adaptabilidade e versatilidade da Etnofísica como ferramenta pedagógica.

Percebemos uma tendência entre os autores citados: a valorização dos saberes ligados à Física, observados em distintos contextos humanos e suas interações com a Física acadêmica. Tal observação nos leva à uma analogia com os modos tradicionais de produção de conhecimento, nos quais os saberes específicos se conectam formando uma "metarracionalidade" abrangente.

Assim, reconhecemos que a Física pode estabelecer diálogos interdisciplinares, enriquecendo nossa percepção sobre o mundo e nossa relação com a natureza. Sabemos que diferentes culturas desenvolveram maneiras distintas de entender fenômenos naturais, os quais têm o potencial de integrar-se à educação em Ciências e Matemáticas (ROSARIO *et al.*, 2018).

Por isso, asseveramos ser crucial examinar as interações entre os conhecimentos físicos e os saberes tradicionais derivados das experiências singulares de cada comunidade, pois ambos se complementam, formando um conjunto mais abrangente, presente em diversas sociedades espalhadas pelo planeta.

Para compreender adequadamente essas interações, o método etnográfico faz-se essencial, exigindo tempo e dedicação para se delinear uma descrição densa e pormenorizada do contexto sociocultural. Em nossa investigação, focamos nos saberes termodinâmicos relacionados à cerâmica, no entanto, reconhecemos que os ceramistas integram conhecimentos de diversas áreas das Ciências em sua prática.

## SABERES FÍSICOS QUE EMERGEM DA PRODUÇÃO DE CERÂMICAS TRADICIONAIS

Situado no estado do Pará, no interior do bioma amazônico brasileiro, o município de Bragança é caracterizado por uma expressiva biodiversidade, evidenciada pelos seus vastos manguezais e igarapés. Em resposta à necessidade de conservação ecológica, instituiu-se, em 2005, a Reserva Extrativista Marinha Caeté-Taperaçu (BRASIL, 2005). Nessa reserva, encontra-se a Vila Cuera, localizada estrategicamente às margens do rio Caeté, a aproximadamente oito quilômetros distante do núcleo urbano de Bragança.

Historicamente, o local, inicialmente ocupado por sociedades indígenas pertencentes à nação tupinambá, configura-se como a origem de Bragança. O termo "Cuera" remete a elementos ancestrais, a algo antigo, do passado, um espaço antigo, consolidando-se na atualidade com dualidades nominativas: "Vila Que Era" Bragança e Vila Cuera. A comunida-de, marcada por tradições, como a produção de cerâmica tradicional, reflete a confluência de culturas indígenas e colonizadoras e a profunda ligação do povo com a terra (ROSARIO, 2018, 2023).

A cerâmica da comunidade Vila Cuera é uma tradição mantida pela família Furtado[3], que amalgama saberes ancestrais transmitidos ao longo das gerações com inovações oriundas de suas próprias experimentações. O processo cerâmico tradicionalmente engloba: a seleção da argila, modelagem, a secagem e a queima da peça. Embora haja nuances importantes em cada fase, neste artigo, daremos ênfase principalmente às fases de modelagem e queima da cerâmica.

A argila empregada na confecção cerâmica da Vila Cuera é extraída das margens do rio Caeté, curso d'água de reconhecida relevância geográfica e cultural para a região[4]. As coletas, geralmente semestrais, se adaptam às necessidades dos ceramistas, visando à otimização e reutilização integral dos materiais, incluindo fragmentos de peças danificadas. Os artesãos locais empregam técnicas refinadas de seleção dessa matéria-prima, utilizando parâmetros visuais, auditivos, táteis e gustativos.

Tal prática originou o termo "cerâmica caeteuara", que representa o vínculo intrínseco entre os ceramistas e o rio Caeté. A extração da argila, frequentemente realizada em áreas acessíveis somente por canoa, ocorre de forma rotativa, sustentando a perpetuação responsável do ofício. A cerâmica caeteuara, em sua morfologia, atua como um registro material da história da região e dos seus primórdios. O método artesanal de produção, aliado à transmissão oral de conhecimentos, exemplifica a interação harmônica entre tradições ancestrais e técnicas contemporâneas.

De acordo com Almeida (2017) e Farias (2006), o legado oral antecede e embasa a racionalidade científica, estabelecendo uma intersecção entre os saberes acadêmicos e tradicionais. No entanto, essa rica herança ceramista enfrenta riscos de declínio, sendo atualmente preservada primordialmente pela família Furtado na Vila Cuera.

Em nossa interação com os artesãos, observou-se a implementação do caraipé[5] durante a preparação da massa argilosa (ROSARIO; SILVA, 2023b), um aditivo frequente nas cerâmicas arqueológicas de diversos povos originários da região amazônica. Isso sinaliza que os protocolos técnicos concebidos por civilizações indígenas pregressas foram, de alguma forma, partilhados ou transmitidos. Prova disso é a presença do caraipé em cerâmicas arqueológicas de distintas regiões paraenses e até de outros estados brasileiros (GUAPINDAIA, 1993; HEPP, 2021; SCHAAN, 2009; SIMÕES, 1981), substância que permanece em uso pelos ceramistas de Vila Cuera.

---

[3] Neste estudo, contou-se com a participação dos ceramistas remanescentes da Vila Cuera, Dona Maria Furtado (mãe) e Josias Furtado (filho), detentores das técnicas tradicionais associadas à produção cerâmica.

[4] O rio Caeté tem diferentes importâncias para o estado do Pará, entre elas, histórica e geográfica, pois foi por esse rio que, há mais de 400 anos, ingleses, franceses e portugueses chegaram ao município de Bragança. Ele também dá origem ao nome da Região de Integração Rio Caeté. Situa-se na mesorregião do Nordeste paraense e compreende parte das microrregiões do Salgado e Bragantina (IOEPA, 2015).

[5] Originário das cinzas da casca e entrecasca de plantas do gênero Licania, esse material é adicionado à pasta de argila, alterando sua textura, porosidade e resistência (ROSARIO; SILVA, 2023b). Também é conhecido como caraipé, caraiperana, caripé, mariperana, milho-torrado-mirim, uxi-do-igapó, uxiranaa (SERVIÇO FLORESTAL BRASILEIRO, 2023). Na região bragantina, é historicamente conhecido como caripé.

No estudo das propriedades físicas da cerâmica, devemos mencionar a etapa prévia à modelagem, pela qual a argila é imersa em água por um período que pode variar de algumas horas a dias, após análise meticulosa de sua composição e características intrínsecas (imagens 1 e 2). Essa etapa é vital para obter uma textura propícia que otimiza o manuseio subsequente.

**Imagem 1:** Argila em depósito  **Imagem 2:** Argila hidratada

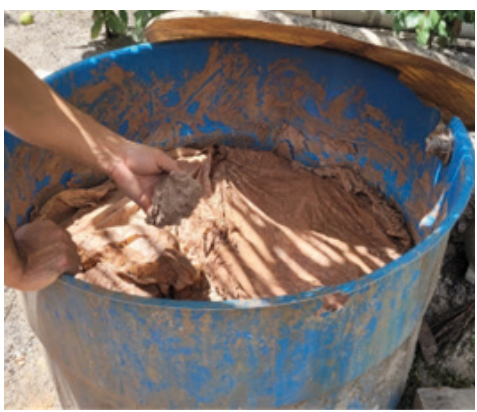 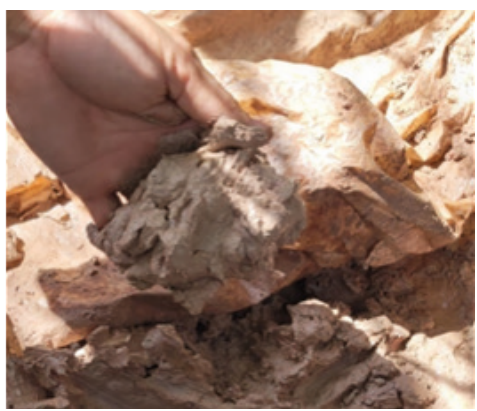

**Fonte:** Acervo da pesquisa, 2021.

Procedimentos rigorosos são observados na formulação da mistura argilosa para modelagem da cerâmica. De acordo com o ceramista Josias Furtado, uma proporção definida de argila, caraipé (imagem 3) e chamote[6] (imagem 4) é essencial[7]. Mediante experimentações rigorosas, estabeleceu-se uma composição ideal para alcançar a consistência da mistura argilosa.

**Imagem 3:** Caraipé triturado e peneirado  **Imagem 4:** Chamote triturado e peneirado

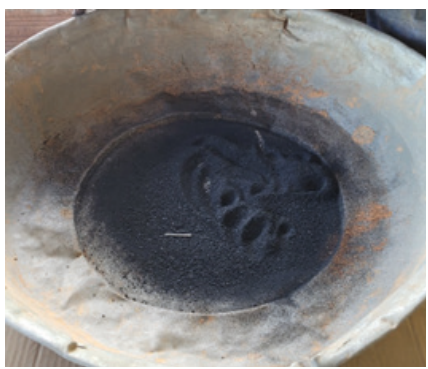 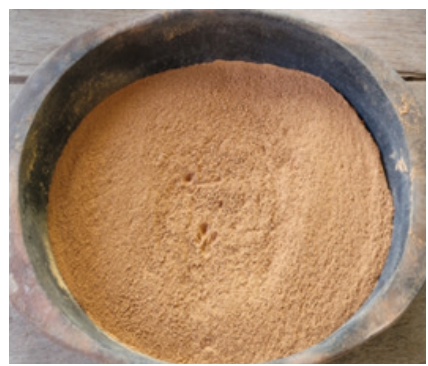

**Fonte:** Acervo da pesquisa, 2021.

---

[6] Fragmentos de cerâmica queimada e triturada ajudam a reduzir a retração durante a queima e aumentam a resistência térmica do corpo cerâmico (ROSARIO; SILVA, 2023b).
[7] Entrevista, dezembro 2022.

Segundo os próprios ceramistas[8], é importante manter uma proporção precisa entre os componentes: para cada três porções de argila, adiciona-se duas de chamote e 0,1 da quantidade de chamote em caraipé. Tal relação é fundamental para atingir a textura e a resistência adequadas, permitindo uma modelagem precisa, resultando em peças cerâmicas de qualidade.

Josias Furtado[9] enfatizou que, com base em múltiplas experimentações, essa proporção é a mais eficaz. Destacou, ainda, o papel do caraipé e do chamote na estabilização das estruturas cerâmicas, especialmente durante as fases de desidratação e transferência de calor nos processos de secagem e queima.

Na fase de modelagem, a quantidade de água adicionada é crucial, transformando a argila bruta em uma massa maleável. Dona Maria[10] advertiu que o estado físico do material depende da qualidade da argila, evitando a excessiva liquidez ou rigidez. A dinâmica envolvida requer uma abordagem específica para cada peça, bem como um gerenciamento adequado da energia manual aplicada.

Do ponto de vista da Física, durante a modelagem (imagens 5 e 6), há uma transição nos estados da matéria da argila, mediada pela quantidade de água adicionada. Essa interação determina a espessura final da peça, influenciando diretamente nos processos termodinâmicos subsequentes, como transferência de calor por condução, convecção e radiação durante a secagem e queima da peça produzida. A espessura determina, em parte, a resistência ao choque térmico, impactando o tempo de secagem e a retração térmica.

**Imagem 5:** Determinando a espessura de um vaso   **Imagem 6:** Determinando a espessura de um prato

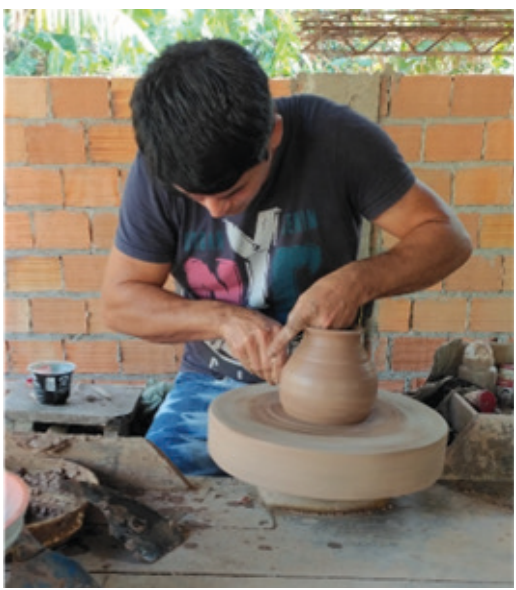 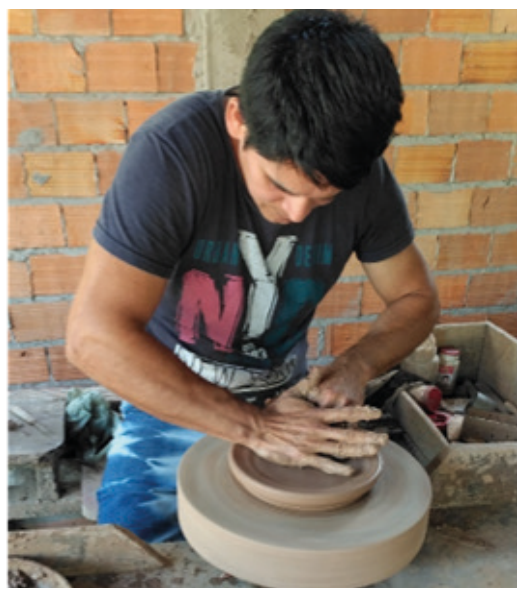

**Fonte:** Acervo da pesquisa, 2021.

---

[8] Informações colhidas por meio de entrevistas e diálogos com Josias Furtado e Dona Maria entre os anos 2021 e 2023.
[9] Entrevista, dezembro 2022.
[10] Entrevista, dezembro 2021.

Após a modelagem, a peça é submetida a uma secagem ao ar livre (imagem 7). Durante esse processo, a evaporação da água induz a contração da peça. A taxa de secagem, influenciada pela espessura, é um parâmetro essencial para prevenir imperfeições, como rachaduras e deformações, provenientes da retração térmica.

**Imagem 7:** Peças em processo de secagem com espessuras e formas variadas.

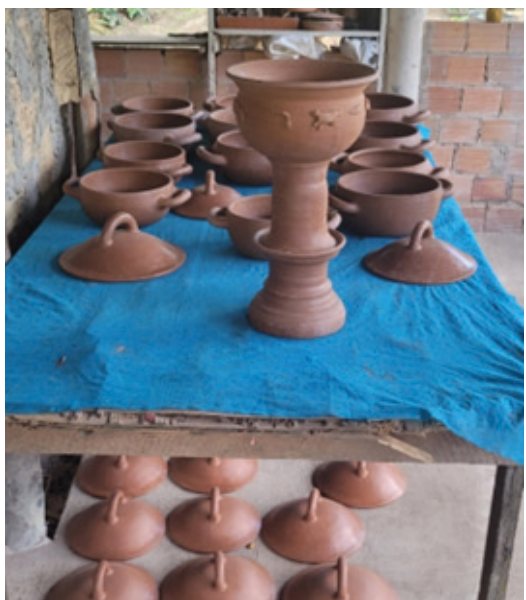

**Fonte:** Acervo da pesquisa, 2023.

A compreensão das interações entre modelagem, espessura e retração térmica é fundamental para produzir uma cerâmica de alta qualidade. Os ceramistas da Vila Cuera, com sua abordagem detalhada e adaptativa, exemplificam a aplicação desses princípios, maximizando a durabilidade e desempenho das peças.

Nessa perspectiva, as habilidades e expertise dos ceramistas da Vila Cuera são cruciais para a produção de uma cerâmica de excelência. Josias e Dona Maria exemplificam a maestria no ofício ao considerar certas variáveis, como os materiais utilizados (argila, chamote e caraipé) e o controle da espessura das peças. A gestão adequada das transferências de calor durante as etapas de produção da cerâmica caeteuara é indispensável para garantir a integridade e a longevidade das peças.

Ao transitarmos da compreensão teórica à aplicação prática, é revelador o testemunho de Josias Furtado e Dona Maria. Durante as entrevistas e as observações, Josias elucidou as dificuldades iniciais encontradas na produção cerâmica. A aprendizagem do ofício, transmitida por sua mãe, dona Maria, e sua avó foi marcada por perdas significativas de peças durante os períodos de secagem e queima. As palavras de Josias ressoam as adversidades enfrentadas:

> Nós tínhamos dificuldade nessa queima, porque a gente não dominava o fogo do forno, queimava as panelas só na fogueira. Aí, eu consegui ajustar esse tempo de queima, essa secagem, que era nossa dificuldade. Então é isso, conseguimos chegar a quase 100%. Porque antigamente perdíamos muitas peças na secagem e na queima e era só prejuízo (Entrevista, julho 2019).

A narrativa de Josias não só ilustra as dificuldades inerentes ao ofício ceramista, mas também a resiliência e a inovação. Por meio de contínuas experimentações, ele e sua mãe melhoraram os métodos de produção, adaptando-se às limitações e transformando desafios em oportunidades de aprendizado. Isso resultou em técnicas refinadas, que hoje garantem uma melhor qualidade do produto final.

Após a fase de desidratação (secagem ao ar livre), inicia-se o processo de sinterização (queima) da cerâmica caeteuara. Josias menciona que, por meio de experimentações sistemáticas, "elevou a eficiência de seu processo". Historicamente, o prolongado tempo no forno indicava uma imprecisão na quantificação da cinética térmica das peças. Contudo, com o pré-aquecimento do forno por longos períodos, Josias notou que a homogeneidade térmica era mais facilmente alcançada, resultando em economia de tempo e controle acurado da temperatura, essencialmente ao manter a combustão constante[11].

Em suas observações diárias, Josias se dedicou à compreensão e à manipulação de fenômenos físicos para otimizar sua produção. Utilizando métodos empíricos e os sentidos humanos, descritos por Lévi-Strauss (2012), estabeleceu parâmetros para a sua prática. Tais princípios – como os conceitos de calor, temperatura, e equilíbrio térmico –, apesar de intrínsecos, não eram compreendidos por meio da linguagem da Física acadêmica.

Uma abordagem sistemática revelou a compreensão do ceramista sobre a cinética térmica no forno: o calor, advindo da parte inferior, atinge inicialmente peças mais densas, enquanto peças com menor espessura, dispostas na parte superior, atingem a temperatura desejada numa etapa seguinte (imagem 8).

Josias fala do design exclusivo de seu forno, cuja arquitetura permite distintas modalidades de transferência de calor, resultando numa sinterização homogênea (imagem 9). A otimização do espaço interno, semelhante à uma técnica de empacotamento, é crucial para maximizar a eficiência e a integridade das peças (diário de campo, julho 2022).

---

[11] Dados obtidos através da observação participante e de entrevistas realizadas entre os anos 2021 e 2023.

**Imagem 8:** Organização das peças no forno   **Imagem 9:** Josias e seu forno

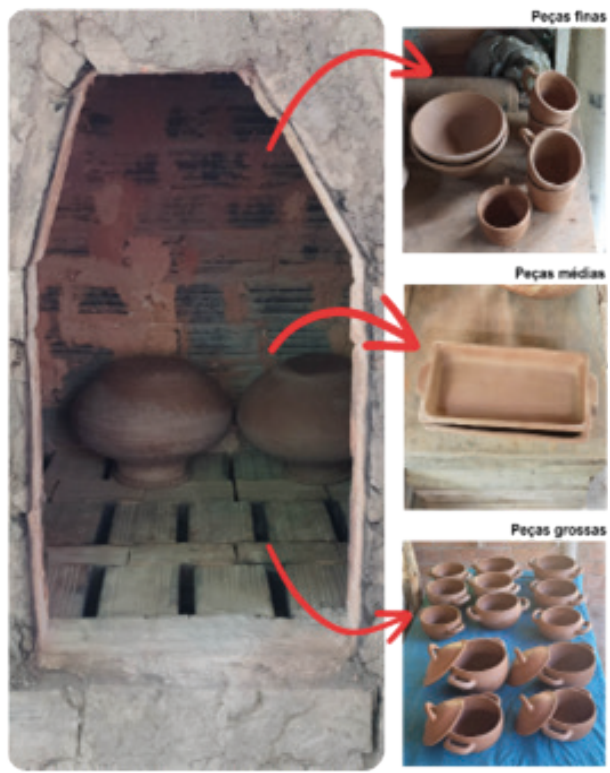 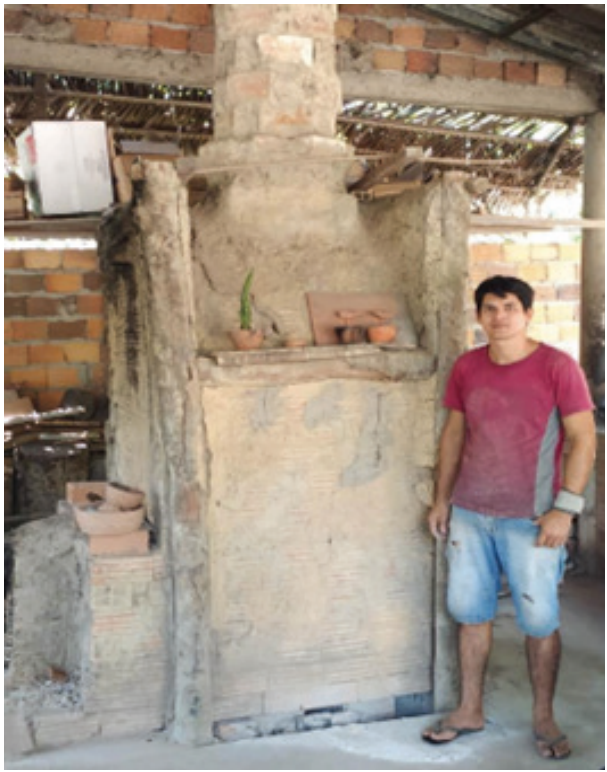

**Fonte:** Acervo da pesquisa, 2023.

A retração térmica, uma consequência inevitável do processo de sinterização, exige uma meticulosa disposição das peças no interior do forno, considerando retrações lineares, superficiais e volumétricas. Josias assegurou que:

> Na hora de arrumar, não pode ter pressa, é no jeito mesmo [...]. A gente vai olhando, vai colocando algumas peças, tirando outras, vendo onde cada peça cabe melhor, deixando um espaço para a peça respirar e comprimir quando ela precisa, [...] a argila quando está queimando diminui de tamanho e depois ela pode aumentar um pouquinho quando está esfriando. Isso tudo é devagar, não é na pressa [..], eu demorei para aprender isso, mas com o tempo a gente aprende e hoje quase não perco mais peça, pois já sei o cuidado que cada peça precisa (Entrevista, dezembro 2022).

Para contrabalancear a retração térmica, várias estratégias são implementadas. Entre elas, são essenciais o uso de suportes termicamente resistentes e um monitoramento empírico constante da temperatura, através da observação da combustão e de emissões gasosas.

Em uma investigação realizada em janeiro de 2023, procuramos entender a eficiência do forno construído por Josias. Uma abertura na porta do forno permitiu a visualização interna durante a sinterização, fornecendo uma perspectiva inédita (imagens 10 e 11).

**Imagem 10:** Porta de vedação com abertura

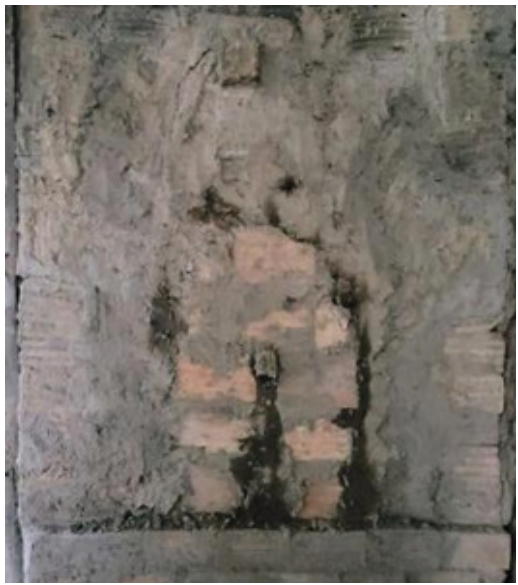

**Imagem 11:** Peças dentro do forno durante a queima

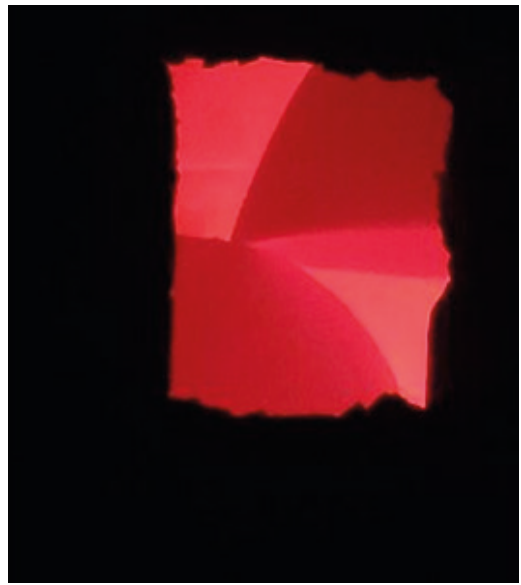

**Fonte:** Acervo da pesquisa, 2023.

É perceptível que os ceramistas, ao longo do tempo, desenvolveram uma compreensão profunda dos princípios da transferência térmica para otimizar seu ofício. O arranjo interno do forno facilita as transferências de calor através dos processos de condução, convecção e radiação; um entendimento que está em sintonia com os fundamentos da Física Térmica. Josias destaca que:

> O tempo no forno é de acordo com a peça. Olho a espessura dela e já tenho uma noção de quanto tempo ela vai ficar lá. Porque o fogo vem de baixo para cima e, aí, o fogo começa a entrar em contato com as peças, vai esquentando o material que está mais perto do fogo. Depois, vai passando a quentura para as outras peças que estão em cima, até o ponto em que as peças ficam na mesma temperatura (Diário de campo, julho 2019).

Por meio dessa descrição, Josias elucida o comportamento energético dentro do forno. Ele ressalta a primazia da condução térmica, pela qual os sólidos transferem calor através de movimentações moleculares. Dentro dessa matriz térmica, o ceramista, alinhando-se com a termodinâmica, manipula e controla variáveis associadas à transferência de calor. Seu método reflete conceitos de condução, convecção e radiação. Essa expertise, edificada sobre tradições e observações empíricas, assegura a eficiência no processo cerâmico.

Os artesãos, intuitivamente, adotam lógicas inerentes à física térmica na confecção de suas peças. Eles aplicam conhecimentos sobre transferência de calor, expansão térmica e princípios de temperatura para metamorfosear a argila em artefatos cerâmicos. Guiando-se por sentidos afinados e por experiências acumuladas, o ceramista determina a energia adequada e a disposição das peças no forno.

Esse *corpus* de conhecimento, todavia, ultrapassa os domínios da termodinâmica. Uma análise acurada do processo cerâmico revela uma gama de princípios físicos mobilizados pelos artesãos. Ao manipular a argila durante a modelagem e controlar a queima das peças produzidas, demonstram um domínio que abrange a mecânica dos materiais e as interações intermoleculares.

Adicionalmente, ao confeccionarem as peças, os artesãos incorporam princípios de ótica e acústica, explorando propriedades reflexivas e sonoras dos artefatos cerâmicos. Tais competências, gestadas empiricamente ao longo de gerações, são evidenciadas nas peças que exibem funcionalidade, resistência e estética refinada.

O meticuloso processo de produção da cerâmica em Vila Cuera destaca uma multifacetada aplicação do conhecimento físico. A maestria com que esses "cientistas da tradição" aplicam e adaptam princípios científicos é um testemunho eloquente da profundidade e da sofisticação de suas tradições ancestrais, sublinhando sua relevância no panorama científico e cultural contemporâneo.

## CONSIDERAÇÕES FINAIS

Ao longo da nossa análise das técnicas empregadas pelos ceramistas da Vila Cuera, é incontestável o profundo entendimento que esses ceramistas detêm sobre os princípios da Física. A forma intuitiva e eficaz com que manipulam a transferência de calor, junto com uma compreensão intrínseca dos conceitos de termodinâmica, serve como testemunho de uma aplicação prática dos princípios que, em ambientes acadêmicos, são estudados de outras formas.

Esse entrelaçamento entre uma prática sociocultural milenar e a Física acadêmica nos conduz ao campo da Etnofísica, que investiga o entendimento e a interpretação dos fenômenos físicos em diversas culturas. A Etnofísica sugere que, muito antes da formalização acadêmica das ciências, as sociedades tradicionais já observavam, entendiam e interagiam com os fenômenos naturais de maneiras que se alinham com os princípios físicos atualmente reconhecidos. Na Vila Cuera, os ceramistas demonstram a aplicação empírica de conceitos da termodinâmica. Essa observação reforça que diferentes culturas desenvolveram suas próprias "ciências", fundamentadas em observações empíricas, tradições e sabedoria transmitidas ao longo de gerações; uma ciência primeira, como lembra Claude Lévi-Strauss (2012).

Mesmo sem a estrutura formal da educação científica, os ceramistas tradicionais têm, ao longo de gerações, cultivado e transmitido um conhecimento que se encontra em surpreendente harmonia com os conceitos fundamentais da Física. Apesar de seus métodos serem embasados em observações empíricas e na sabedoria ancestral, os resultados que alcançam dialogam diretamente com as aspirações e os entendimentos da ciência.

Neste estudo, ao buscarmos elucidar as interseções entre a Física e os saberes da tradição, emergiu uma clara congruência entre os dois domínios. Na superfície, podem parecer universos apartados, devido às suas diferenças de linguagem, de métodos e de abordagem. Contudo, um olhar perspicaz revela que eles estão, de fato, entrelaçados. Os ceramistas, por meio da aplicação implícita de princípios físicos em sua arte, proporcionam uma perspectiva que não apenas complementa, mas também enriquece a Física tal como entendida nos moldes acadêmicos.

Esse entrelaçamento entre a ciência contemporânea e as tradições ancestrais destaca-se como um testemunho da colaboração e da complementaridade entre ambas. A Física, em seu núcleo, beneficia-se das intuições e *insights* fornecidos por essas tradições, assim como as práticas tradicionais encontram paralelismo e ressonância nos princípios da Física acadêmica. Portanto, ao reconhecer e valorizar os saberes da tradição, reafirmamos não só a relevância das tradições em nosso mundo contemporâneo, mas também ressaltamos a contínua e harmoniosa convergência entre ciência moderna e tradição.

## REFERÊNCIAS

## FONTES ORAIS

Josias Furtado Padilha.

Maria de Nazaré Furtado da Silva.